\documentclass[12pt,preprint]{aastex} 

\begin{document} 

\title{{\it{FUSE}} Observations of the Dwarf Novae UU Aql, BV 
Cen, and CH UMa in Quiescence\altaffilmark{1}} 

\author{Edward M. Sion, Patrick Godon\altaffilmark{2}, 
Fuhua Cheng} 
\affil{Astronomy and Astrophysics, Villanova University, \\ 
800 Lancaster Avenue, Villanova, PA 19085, USA} 
\email{edward.sion@villanova.edu, 
patrick.godon@villanova.edu, 
fcheng@ast.vill.edu} 

\author{Paula Szkody} 
\affil{Department of Astronomy, 
University of Washington, 
Seattle, WA 98195, USA} 
\email{szkody@astro.washington.edu} 

\altaffiltext{1}
{Based on observations made with the NASA-CNES-CSA Far Ultraviolet
Spectroscopic
Explorer. {\it{FUSE}} is operated for NASA by the Johns
Hopkins University under NASA contract NAS5-32985}
\altaffiltext{2}{Visiting at the 
Space Telescope Science Institute, Baltimore, MD 21218, USA; godon@stsci.edu} 

%\clearpage 

\begin{abstract} 

We report on {\it{FUSE}} spectra of three U Gem-type, long period, 
dwarf novae, 
UU Aql, BV Cen and CH UMa taken during their quiescence intervals. We 
discuss the line identifications in their spectra and attempt to 
characterize the source(s) of their FUV flux distribution. 
Archival {\it{IUE}} spectrum of CH UMa and BV Cen in quiescence 
were identified as having a matching flux level with the {\it{FUSE}} spectra 
and these were combined with each {\it{FUSE}} spectrum 
to broaden the wavelength coverage and further 
constrain model fits. Multi-component synthetic spectral fits from our 
model grids, consisting of single temperature white dwarfs, 
two-temperature white dwarfs, accretion disks and white dwarfs plus 
accretion disks, were applied to the {\it{FUSE}} spectra alone and to the 
combined {\it{FUSE}} + {\it{IUE}} spectra. We present the results of 
our model analyses and their implications. 

\end{abstract} 


\section{Introduction} 

Dwarf novae (DNe) are close interacting binaries in which a Roche-lobe 
filling main sequence-like dwarf transfers matter with angular momentum 
through a disk onto a white dwarf (WD). The rapid disk accretion during 
outburst, due to a thermal instability that causes cyclic changes of the 
accretion rate, releases gravitational potential energy identified as the 
DN outburst.  The high accretion rate ($\sim 10^{-8}$ to $10^{-9}$ 
$M_{\odot}$/yr) outburst phase (which lasts a few days to weeks) is 
preceded and followed by a low accretion rate ($\sim 10^{-11} 
M_{\odot}$/yr)  quiescence stage. This DN behavior is believed to be 
punctuated every few 
thousand years or more by episodes of catastrophically unstable 
thermonuclear burning, the classical nova explosion.  Perhaps the least 
understood topic in CV/DN research (along with what drives the wind 
outflow in outburst) is the state and structure of the boundary layer and 
accretion disk during quiescence and the physics of how long term 
accretion of mass, angular momentum and energy affects the WD.  Our 
studies with archival {\it{IUE}}, and {\it{HST}}/STIS have found 
that $\sim$50\% of the 
DNe in quiescence are dominated (i.e., $>$ 60\% of UV flux) by a 
component of FUV
flux other than the WD called the "accretion disk"; $\sim$25\% are dominated 
by the WD and $\sim$25\% have nearly equal contribution of WD and accretion 
disk (40-60\% each) \citep{urb06}.  

A number of studies \citep{ara03,sio91,sio99,urb06} 
have shown that CV WDs above the gap are typically 
on-average ~ 10,000K hotter than CV WDs below the period gap (almost 
certainly a consequence of higher time-averaged accretion rates of systems 
above the gap but possibly with system total age also being a factor).  
Since the white dwarf surface temperature is crucial for understanding CV 
evolution and whether CVs evolve across the period gap, the use of 
cooling ages and long term evolutionary model sequences with accretion 
(including the effects of nova explosions, \citet{tow03}) 
must rely on the empirical WD temperature of the photosphere in 
equilibrium with 
long term compressional heating from accretion. The work of 
\citet{tow02,tow03} allows measured T$_{eff}$'s of CV WDs to be 
converted to the accretion rate per unit WD surface area averaged over the 
thermal time of the WD envelope. 

Unfortunately, there are far fewer systems with reliably known WD 
properties above the period gap compared with below the gap, 
thus impeding detailed comparisons between the two groups. For example, 
among CVs below the gap, there are now roughly 20 systems with reliable WD 
temperatures but only 5 systems above the gap with reliable WD 
temperatures. The primary reason for this disparity is that in long period 
CVs with higher mass transfer rates, the disks may remain optically thick 
even during quiescence, making the disk contribution to the total flux 
typically larger in systems above the gap. Hence, it is more difficult to 
disentangle the white dwarf flux contribution from that of the accretion 
disk. 

As part of our effort to increase the sample of CV degenerates with known 
properties above the gap, we have used {\it{FUSE}} 
and {\it{IUE}} archival spectra to 
analyze three long period dwarf novae, UU Aql, BV Cen, and CH UMa. For UU 
Aql, system properties were adopted from \citet{rit03} 
and from \citet{szk87}. 
For BV Cen parameters were adopted from \citet{rit03}. 
For CH UMa, we adopted values from \citet{fri90}. 
For all three systems, the distances were the same as those in 
\citet{urb06} where the \citet{war95} and \citet{har04} 
M$_{v(max)}$ versus P$_{orb}$ relations, calibrated with trigonometric 
parallaxes, were used. The reddening values were the same as those quoted 
in \citet{urb06} which were from \citet{ver87}, \citet{lad91} 
and \citet{bru94}. 

The dwarf nova systems analyzed in this work are UU Aql, BV Cen, and CH 
UMa. In Table 1, the observed properties of these dwarf novae are 
summarized by column as follows: (1) system name; (2) dwarf nova subclass 
with UG denoting a U Gem-type system; (3) 
orbital period in days; (4) the recurrence time of dwarf nova outbursts in 
days; (6) the apparent 
magnitude at minimum (quiescence); (7)  the apparent magnitude in 
outburst; (8) secondary spectral type; (9) orbital inclination in degrees; 
(10) white dwarf mass in solar masses; (11) secondary star mass in solar 
masses; (12) adopted reddening value and ; (13) distance in parsecs. 

\section{Observations and Data Reduction} 

The instrumental setup and exposure details of 
the {\it{FUSE}} spectra of BV Cen, 
CH UMa and UU Aql in quiescence are provided in Table 2. The LWRS was used 
in all cases since it is least prone to slit losses due to the 
misalignment of the four {\it{FUSE}} telescopes.
All the spectra were obtained in time tag (TTAG) mode, and each one of them  
consists of 7 individual exposures (corresponding to 7 {\it{FUSE}} orbits). 
It is clear that the relatively 
poor {\it{FUSE}} spectral quality of the spectra speaks to the requirement for 
more observing time. Nevertheless, we deemed that there was sufficient S/N 
to warrant a first attempt multi-component FUV analysis of each system 

All the data were reduced using CalFUSE version 3.0.7.  
In this version of CalFUSE the data are maintained as a photon 
list: the intermediate data file - IDF.
Bad photons are flagged but not discarded, so the user can examine
and combined data without re-running CalFUSE. 
For each target, we combined the individual exposures (using the IDF
files) and channels to create a time-averaged spectrum 
weighted in the flux in each output datum by the exposure time 
and sensitivity of the input exposure and channel of origin. The details 
are given here. The spectral regions 
covered by the spectral channels overlap, and these overlap regions are 
then used to renormalize the spectra in the SiC1, LiF2, and SiC2 channels 
to the flux in the LiF1 channel. We then produce a final spectrum that 
covers almost the full {\it{FUSE}} wavelength range $905-1182$\AA. The low 
sensitivity portions of each channel are discarded. In most channels there 
exists a narrow dark stripe of decreased flux in the spectra running in 
the dispersion direction. This stripe has been affectionately known as the 
''worm'' and it can attenuate as much as 50\% of the incident light in 
the affected portions of the spectrum. The worm has been observed to move 
as much as 2000 pixels during a single orbit in which the target was 
stationary. The "worm" appears to be present in every exposure and, at 
this time, there is no explanation for it. Because of the temporal changes 
in the strength and position of the worm, CalFUSE cannot correct target 
fluxes for its presence. Here we take particular care to discard the 
portion of the spectrum where the so-called \textit{worm} 'crawls', which 
deteriorates LiF1 longward of 1125\AA\ . Because of this the $1182 - 
1187$\AA\ region is lost. 
We then rescale and combine the spectra. When we combine, we weight 
according to the area and exposure time for that channel and then rebin 
onto a common wavelength scale with a 
$0.1$\AA ,  $0.2$\AA , and  $0.5$\AA\  resolution. 

In the observing log given in Table 2, the entries are by column: (1) 
gives the target, (2) {\it{FUSE}} spectral data ID, 
(3) the aperture used, (4) the 
date and time of observation, (5) the (good) 
exposure time in seconds, (6) central wavelength,
and (7) S/N. 

The {\it{FUSE}} spectra for the three systems, UU Aql, BV Cen and CH Uma are 
displayed in Figures 1, 2, and 3 respectively. 
A quantitative
sense of the relative data quality is provided by the signal to noise
for the three {\it{FUSE}} spectra. We binned the data by 0.1\AA\ for which 
the S/N of UU Aql, BV Cen, and CH UMa is 5.15, 5.9, and 3.6, respectively.    

For UU Aql, Fig. 1 reveals 
a rich line spectrum with numerous lines of molecular hydrogen, 
interstellar species and possible accretion disk or photospheric 
absorption features. The spectrum reveals a downturn in the continuum 
shortward of 1000\AA . The spectrum does not exhibit any evidence of emission 
lines from the source, the only emission lines are from air glow and 
heliocoronal (e.g. sharp emissions lines from 
C\,{\sc iii} (around 977\AA ) and the O\,{\sc vi} doublet). 

In figure 2, the {\it{FUSE}} spectrum of BV Cen has a variety 
of interstellar and stellar features but has a continuum shape distinctly 
different from UU Aql. The broad C\,{\sc iii} 
absorption feature around 1175\AA\ 
is definitely from the source, and the C\,{\sc iii} (around 977\AA ) 
and the O\,{\sc vi} doublet broad emission features could also possibly be   
associated with one of the FUV components in BV Cen. 
All the other sharp emission 
features are either heliocoronal or geocoronal in origin.   

In figure 3, CH UMa reveals numerous absorption 
lines due to highly ionized and singly ionized metals. It has a continuum 
energy distribution similar to UU Aql. There are definitely some broad emission 
lines from the source itself. The most prominent one is the O\,{\sc vi} 
doublet (the right component being strongly 
attenuated by molecular hydrogen sharp absorption 
lines), C\,{\sc iii} (both around 977\AA\ and 1175\AA) which also 
seems to be in emission and  
a tentative identification of N\,{\sc iv} 
emission in the short wavelengths. The source is 
also contaminated with sharp emission lines due to air glow. Here too the 
N\,{\sc i} \& N\,{\sc ii}  
are geocoronal in origin, and the sharp peaks on top of 
the broad C\,{\sc iii} (977\AA) and O\,{\sc vi} 
emissions are heliocoronal in origin.    
We note that ISM molecular hydrogen absorption 
is affecting the continuum.   

Since the wavelength range covered by {\it{FUSE}} 
overlaps with {\it{HST}}/STIS or {\it{IUE}} 
in the region of C\,{\sc iii} 
between 1170\AA\ and 1180\AA, a much broader FUV 
wavelength coverage is afforded by combining the spectra when the flux 
levels of the two spectra in the wavelength overlap region match closely 
enough. We found archival {\it{IUE}} spectra matching the {\it{FUSE}} 
spectra of two
of the three systems (CH UMa and BV Cen) but unfortunately no {\it{HST}} 
spectra exist for the three systems. 

The {\it{FUSE}} + {\it{IUE}} combination of spectra 
rests on the assumptions that (1) 
differences between the two spectra in orbital phase and (2) in time 
after the last outburst can be ignored. Given the long exposure times of
the {\it{FUSE}} and {\it{IUE}} spectra, the questionable 
reliability of the orbital ephemerides, including UU Aql's and (3) the limited 
S/N of the quiescent spectra, the influence of phase-dependent variations 
is not considered. However, the time since the last outburst as well 
as the brightness state of the system at the 
times of the {\it{FUSE}} and {\it{IUE}} 
observations for the three systems is considered in detail using AAVSO 
archival light curve data. For UU Aql, the {\it{FUSE}} spectrum was obtained 
approximately 50 days after its last major dwarf nova outburst,
however, the {\it{IUE}} spectrum was obtained during the transition 
to a brightening that appeared not to be a major outburst. Therefore, 
we have excluded the {\it{IUE}} spectrum from use in combination with the 
{\it{FUSE}} data for UU Aql. For BV Cen both 
the {\it{FUSE}} and {\it{IUE}} spectra were 
obtained during quiescence but the {\it{FUSE}} 
spectrum was acquired approximately 
159 days after the last outburst while the {\it{IUE}} spectrum was taken 
roughly 50 days after the last outburst. Thus, the {\it{FUSE}} spectrum 
probably recorded a greater degree of white dwarf cooling than the 
{\it{IUE}} spectrum obtained closer to the last outburst. Likewise for CH UMa, 
both the {\it{FUSE}} and {\it{IUE}} spectra 
were obtained during quiescence but the {\it{FUSE}} 
spectrum appears to have been obtained roughly 125 days after the last 
outburst while the {\it{IUE}} spectra were acquired about 83 days after the 
last outburst. Since the e-folding times for white dwarf cooling in 
both of these systems following the outburst 
heating episode is typically shorter than the above two post-outburst 
intervals, it is probably acceptable to 
combine the {\it{FUSE}} and {\it{IUE}} spectra for BV Cen and CH UMa.    

In Table 3 we present the observing log of {\it{IUE}} observations which 
matched the {\it{FUSE}} flux level in the wavelength overlap region of these 
two systems. The entries by column are (1) the target name, 
(2) the observation ID, (3)  aperture, (4) dispersion mode, (5) date of 
the observation, (6) time of mid-exposure, and (6) the exposure time in 
seconds. 

Thus, our analysis was carried out first for 
the {\it{FUSE}} spectra of the three
systems and then separately for the combined 
{\it{FUSE}} plus {\it{IUE}} data of BV Cen
and CH UMa.

\section{Multi-Component Synthetic Spectral Fitting} 

Our data analysis and modeling involves the full 
suite of multi-component (accretion disk, white dwarf photosphere, 
accretion belt)  synthetic spectral codes, 
which we have utilized in 
our spectral fitting of {\it{FUSE}} and {\it{IUE}} data. Based upon our 
expectation that the accreting white dwarf is an important source of FUV 
flux in these systems during quiescence, we carried out a high gravity 
photosphere synthetic spectral analysis first. The model atmosphere 
(TLUSTY200; \citet{hub88}), and spectrum synthesis codes 
(SYNSPEC48 and ROTIN4 \citet{hub95}) and details of 
our $\chi^{2}_{\nu}$ ($\chi^2$ per degree of freedom) minimization fitting 
procedures are discussed in detail in \citet{sio95} and will not be 
repeated here.  To estimate physical parameters, we generally took the 
white dwarf photospheric temperature T$_{eff}$, log $g$, and rotational 
velocity $v_{rot}$ and chemical abundances as free parameters.  

We normalize our fits to 1 solar radius and 1 kiloparsec such that the 
distance of a source is computed from $d = 
1000(pc)*(R_{wd}/R_{\sun})/\sqrt{S}$, or equivalently the scale factor $S 
= \left( \frac{R_{wd}}{R_{\odot}} \right)^{2} \left( \frac{d}{kpc} 
\right)^{-2}$, is the factor by which the theoretical flux (integrated 
over the entire wavelength range) has to be multiplied to equal the observed 
(integrated) flux. 

The grid of WD models extended over the following range of parameters: log $g 
= 7.0, 7.5, 8.0, 8.5, 9.0$; T$_{eff}/1000$ (K) = 22, 23, ..., 75; Si = 
0.1, 0.2, 0.5, 1.0, 2.0, 5.0; C = 0.1, 0.2, 0.5, 1.0, 2.0, 5.0; and 
$v_{rot} \sin{i}$ (km~s$^{-1}$) = 100, 200, 400, 600, 800. 

For the synthetic accretion disk models, we used the latest
accretion disk models from the optically thick disk model grid
of \citet{wad98}. The range of disk model parameters varies as follows:
WD mass (in solar units) value of 0.35, 0.55, 0.80, 1.03, and 1.21;
orbital inclination (in degrees) of 18, 41, 60, 75 and 81.
The accretion rate ranges from $10^{-10.5}M_{\odot}$yr$^{-1}$ to   
$10^{-8.0}M_{\odot}$yr$^{-1}$ by increments of 0.5 in $log{\dot{M}}$.  

For each dwarf nova, we adopted the following procedure. First, we 
masked out all of the obvious emission features 
and artifacts in both the {\it{FUSE}} 
and {\it{IUE}} spectra of each object. 
Second, we carried out synthetic spectral 
fits using and/or combining model components in this order: a white dwarf 
model alone, accretion disk model alone, combination white dwarf plus 
accretion disk model, and two-temperature white dwarf model (the latter to 
simulate a hotter equatorial region as well as a cooler photosphere at 
higher latitudes). For accretion disk fits, we "fine-tuned" the 
derived accretion rate of the best-fitting disk model by changing the 
accretion rate in increments of 0.1 over the range 0.1 to 10, on the 
assumption   
that the disk fluxes scale linearly over that range. 

In Table 4, we indicate where we masked any strong emission features, 
artifacts, or negative fluxes in the {\it{FUSE}} and {\it{IUE}} spectra of each object. 

\section{Synthetic Spectral Fitting Results} 

The noise level of the {\it{FUSE}} spectra precludes the opportunity to extract 
reliable parameters for the accreting white dwarfs in these three systems. 
This is especially true for deriving rotational velocities which rely on 
well-resolved, strong absorption lines arising in the photosphere. The 
rotational velocity is also affected by underlying emission filling of 
absorption features and by the chemical abundances one uses. With these 
caveats in mind, we proceeded to apply our grid of WD photosphere models 
(keeping the chemical abundance fixed at solar) and accretion disk 
synthetic spectra. 

For any dwarf nova in quiescence, a single 
temperature white dwarf model should be 
a reasonable first approximation as the source 
of the FUV flux. For UU Aql, we adopted 
two possible distances, 150 pc and 350 pc and carried out detailed fits 
for both values. For a distance of 350 pc, the best-fit WD model to the 
{\it{FUSE}} spectrum gave T$_{eff} = 27,000$K, R$_{wd}$/R$_{\sun} = 1.13 \times 
10^{-2}$ and a $\chi^{2} = 0.963$. This best-fitting model is shown in 
figure 4. The continuum of the model gives a fair representation of the 
observed continuum down to about 1060\AA\ but there is a shortfall of model 
flux relative to the data at wavelengths shorter than 1030\AA. For the same 
distance, an accretion disk alone yielded a best fit with $\chi^{2} = 
1.14$, an accretion rate of $5\times 10^{-11} M_{\sun}$/yr and inclination 
$i = 41$ degrees, M$_{wd} = 0.8 M_{\sun}$. This disk fit is shown in 
figure 5. The model disk continuum, unlike the WD, fails to match the flux 
level of the data between about 1090\AA\ and 1180\AA\ and the solar abundance 
accretion disk model fails to provide a sharp absorption features. 

A combination white dwarf plus accretion disk yielded a modest improvement 
with $\chi^{2} = 0.71$, $\dot{M} = 1.6 \times 10^{-11} M_{\sun}$/yr, $i = 
41$ degrees and $M_{wd} = 0.8 M_{\sun}$. Finally we tried two temperature 
(WD + belt) fits. The best fit two temperature WD yields $\chi^{2} = 0.74$ 
with the cooler white dwarf portion ( T$_{eff} = 24,000$K) giving 57\% of 
the flux and the hotter belt (T$_{belt} = 33,000$K) providing 43\%
(figure 6).  

In general, for a distance d = 150 pc, the model fits to UU Aql's {\it{FUSE}} 
spectrum are worse than for 350 pc. Applying single temperature white 
dwarf fits, we obtained T$_{eff} = 17,000$K, $R_{wd}/R_{\sun} = 1.06 
\times 10^{-2}$ and a $\chi^{2} = 1.13$. An accretion disk alone yielded a 
best fit with $\chi^{2} = 1.23$, an accretion rate of $1.3 \times 10^{-11} 
M_{\sun}$/yr and inclination $i = 41$ degrees and M$_{wd} = 0.8 M_{\sun}$. 
A combination white dwarf plus accretion disk yielded a modest improvement 
with $\chi^{2} = 1.13$, $\dot{M} = 1.6\times10^{-11} M_{\sun}$/yr, $i = 
41$ degrees and $M_{wd} = 0.8 M_{\sun}$. A two temperature white dwarf (WD 
+ belt) gives a best-fit with a $\chi^{2} = 1.02$, with a 17,000K white 
dwarf providing 77\% of the FUV flux and the belt giving 23\% of the flux.
In view of the much better agreement of the models with the observations of 
UU Aql for our assumed distance of 350 pc than for a distance of 
150 pc, the closer distance can be ruled out.

For CH UMa, the {\it{FUSE}} spectrum is very noisy 
and underexposed. We took into
account our adopted distance of 300 pc in the model fitting, we fixed the
WD mass at M$_{wd} = 1.2 M_{\sun}$ (Log $g = 9.0$, with a radius of
$\approx 4,000$km), and fixed the disk inclination at the published value
of 18 degrees. A single temperature white dwarf fit to CH UMa had
$\chi^{2} = 0.227$, and yielded a best fit T$_{eff} = 29,000$K, and a
distance of 310pc.  This model however did not fit very well in the
shorter wavelengths ($<$1020\AA ).  A lowest $\chi^{2}$ fit for this same WD
mass was obtained for T=40,000K, with $\chi^{2} = 0.208$, but it yielded a
distance of 600pc, twice the adopted estimate of the distance.  Since the
Ritter catalog give a mass of 1.95$M_{\odot}$ (well above the Chandraskhar
mass limit for a WD), we decided to try a larger mass with a
correspondingly smaller radius which lead to a smaller emitting surface
area and therefore a shorter distance. We assumed $M=1.38M_{\odot}$
($log {g} = 9.5$, with a radius of $\approx 2,000$km) and found that the best
fit was for $T=40,000$K. This model yielded a distance of 307pc with
$\chi^{2} = 0.199$. Since the absorption features around 
1120\AA\ - 1150\AA\ are
not pronounced in the observed spectrum, we decided to increase the
rotational velocity to improve the fit of this model. However, a better
result was obtained by simply reducing the abundances of Si and C to 0.01
their solar value.  This low Si and C model had a $\chi^{2} = 0.184$ and a
distance of 314pc. This model is shown in Figure 7. Because of the low
S/N of the spectrum the assessment of the error on the temperature
estimate is of the order of 5,000K, namely $T_{wd}=40,000\pm 5,000$K.

Though the WD fit yielded the lowest value of $\chi^2$, we tried 
accretion disk fits alone to CH UMa's {\it{FUSE}} data. The best fit gave a 
$\chi^{2} = 0.213$, an accretion rate of $5 \times 10^{-12} M_{\sun}$/yr 
for $M_{wd} = 1.2 M_{\sun}$ and $i = 18$ degrees. 

Various attempts to fit the combined {\it{FUSE}} 
+ {\it{IUE}} spectra of CH UMa met with 
limited success as summarized in Table 5. The best combination fit was for an 
accretion disk plus WD with the disk contributing 79\% of the FUV flux. However,
these fits were less satisfactory than the fits 
to the {\it{FUSE}} spectrum alone.

Because the {\it{FUSE}} spectrum of CH UMa is very noisy, and because of the
broad emission lines and the absence of strong absorption lines, it is
difficult to assess which model (disk or WD) is the best solution.
However, the absence of strong absorption lines would favor the 
accretion disk model because of Keplerian broadening.  
Also because of the poor quality of the spectrum, 
composite models (WD+accretion disk, two-temperature WD) 
did not improve the fit. In view of all of the above, we cannot be confident 
that we have determined the temperature of the white dwarf in CH UMa.  

Next we analyzed the combined {\it{FUSE}} + {\it{IUE}} spectra of
CH UMa. 
The best white dwarf-only fit yielded T$_{eff} = 31,000$K, log 
$g = 9$ a $\chi^{2} = 3.02$ and $R_{wd}/R_{\sun} = 6.02 \times 10^{-3}$.   
For accretion disk models only, the best fit occurred with a $\chi^{2} = 
2.02$, corresponding to $\dot{M} = 3 \times 10^{-9} M_{\sun}$/yr. This 
disk fit is a modest improvement of the fit to CH UMa since the $\chi^{2}$ 
value was lowered to 2.02. The best-fitting white dwarf + accretion disk 
model  resulted in only a modest, statistically 
insignificant improvement. The $\chi^{2}$ value was lowered to 1.87, the 
accretion rate was $6.4 \times 10^{-11} M_{\sun}$/yr with the WD T$_{eff} 
= 22,000$K and the scale factor yielding a white dwarf radius of $5.84 
\times 10^{-3} R_{\sun}$. In this composite disk plus white dwarf fit, the 
WD contributes 21\% of the FUV flux while the accretion disk contributes 
79\% of the FUV flux. We also tried a two-temperature 
WD solutions with the best-fitting model consisting of a 26,000K WD 
providing 77\% of the flux and a hot accretion belt/ring with T$_{belt} = 
50,000$K giving 23\% of the UV flux. However, this two-temperature fit was 
no better than the WD + accretion disk fit. 

For BV Cen we dereddened the spectrum assuming E(B-V)=0.10, 
and first tried single temperature white dwarf fits with a white 
dwarf mass $= 0.83 M_{\sun}$, and 
used solar chemical abundances. We found the best-fitting white dwarf 
model to have T$_{eff} = 40,000\pm 1000$K, log $g= 8.3$, $V \sin i = 500$ 
km/s $\pm 100$ km/s. This fit yielded $\chi^{2} = 0.2701$ and 
a distance d=435pc (see figure 8). 
We note here that the WD solution
fits the following absorption features quite accurately: 
C\,{\sc ii} (1066\AA ), 
S\,{\sc iv} (1073\AA ), 
N\,{\sc ii} (1084\AA ), 
S\,i{\sc iv} (1122.5\AA , 1128.3\AA ), and  
S\,i{\sc iii} (1140\AA - 1146\AA ). 

For an optically thick, steady state accretion disk alone, 
we chose  $M_{wd} = 0.80 M_{\sun}$, $i = 60 $ degrees.
The best fit we obtained has a mass accretion rate
of $10^{-8.5}M_{\odot}$yr$^{-1}$, too large for dwarf nova quiescence. 
Moreover, the longer wavelength part of the spectrum
is rather flat and unable to fit the absorption features around
1120\AA\ and 1130\AA. This fit, with i=60deg,  leads to a distance
of 1255pc. In order to fit these absorption features, one needs to assume an
inclination of 18 degrees, inconsistent with the known inclination
of the system (62 degrees). Such a model leads to a distance of
more than 2000pc, again inconsistent with all estimates of the system distance 
(500pc). Therefore, based on the parameters of the system,
the disk solution is completely inconsistent. We tried 
composite model fits, but they also led to very poor results.  
We also combined BV Cen's {\it{FUSE}} + {\it{IUE}} spectra but they  
led to very poor fits, much worse than the fits to the {\it{FUSE}} spectrum
alone. Since the WD solution is consistent with the parameters of the 
system and fits the absorption features of the spectra, it is clear that the
favored solution of the {\it{FUSE}} spectrum of 
BV Cen is $T_{wd}=40,000\pm 2,000$K.

\section{Discussion} 

Our principal objective of determining the surface temperatures of the 
white dwarfs in these three dwarf novae during quiescence has met with 
mixed results. From synthetic spectral fits to the {\it{FUSE}} spectra of the
three long period dwarf novae, UU Aql, CH UMa and BV Cen, and the {\it{FUSE}} 
+ {\it{IUE}} archival SWP spectra of CH UMa we have presented preliminary 
evidence that during quiescence, their accreting white dwarfs all have
surface temperatures hotter than 20,000K. Unfortunately, all three 
temperatures have considerable uncertainty due to 
the low S/N of the {\it{FUSE}} spectra and 
{\it{IUE}} spectra as well as the difficulty 
of disentangling the flux contribution of the second component of FUV flux
or "accretion disk" during  quiescence. Of the three systems in this study, 
we regard our estimate of the $T_{eff}$ of the WD in UU Aql to be the most 
reliable since that system appears to be dominated in the FUV by the white 
dwarf flux. 

For UU Aql, we used both assumed distances, 150 pc and 350 pc, and tried 
the composite fits (WD + disk) to UU Aql's {\it{FUSE}} spectrum. However, 
the quality of the fits for WD-only, disk-only, WD + disk, and WD + "belt" 
were all roughly comparable and thus it is difficult to distinguish the 
best-fit case. For both distances, statistically insignificant 
improvements in the fits result when a white dwarf and accretion disk are 
combined or a two-temperature WD (WD + belt) is applied to the {\it{FUSE}}  
data. The results in Table 5 illustrate the difficulty. Qualitatively, the 
composite fits involving combinations of white dwarf plus accretion disk 
or accretion belt models look marginally more reasonable than the fits 
that involve single component (WD or Disk). It appears that for both 
distances, the white dwarf component is the dominant source of FUV flux 
and that the $T_{eff}$ of the WD is probably between 17,000K and 27,000K
(say 22,000$\pm$5,000K). 

For BV Cen, the WD model fit to the {\it{FUSE}} spectrum gave the best result,
both in fitting actual features of the observed spectrum and in leading
to consistent values of the system parameters. This was not the case
for the disk model and the composite models. 
From the best WD model fit we obtained that the
WD of BV Cen must have a temperature of about 40,000K.  

For CH UMa, the broad and dominant emission lines together with
the poor S/N (and possible detector noise at very short wavelengths)
of the {\it{FUSE}} and {\it{IUE}} spectra precluded the opportunity to obtain
conclusive results. However, the results for the {\it{FUSE}} spectrum
were of a much higher quality than for the 
{\it{FUSE}} + {\it{IUE}} combined spectrum. 
Therefore we adopt the {\it{FUSE}} results for CH UMa with the possibility
that the WD temperature could be as high as 40,000K. 
 
In Table 6, we list the dwarf novae above the period gap whose 
white dwarfs have surface temperature determinations. In Column (1) we 
give the system name, column (2) the orbital period; column (3) the surface 
temperature and column (4) the temperature reference. As seen in Table 6, 
there is now a sample of eight long period dwarf novae 
of which roughly seven have relatively secure white dwarf temperatures 
obtained during quiescence. In the case of BV Cen, the 
inclination is expected to be high. Thus, it is plausible to expect that 
the quiescent accretion disk may be blocking the direct radiation from the 
accreting white dwarf.  

The relatively poor {\it{FUSE}} spectral quality of CH UMa 
underscores the need for  re-observation with longer exposure 
times. However, analysis will remain hindered until more reliable 
information on white dwarf masses and distances becomes available. Until 
then, the conclusions in this work must be regarded as preliminary. 

\section{Acknowledgements} 

We thank an anonymous referee for helpful comments and corrections.
PG wishes to thank the Space Telescope Science 
Institute for its kind hospitality. 
This work was supported by NSF grant AST05-07514 and NASA 
grants NAG5-12067 and NNG04GE78 to Villanova University. 
This research was based on observations made with the
NASA-CNES-CSA Far Ultraviolet Spectroscopic Explorer.
{\it{FUSE}} is operated for NASA by the Johns Hopkins University under
NASA contract NAS5-32985.

\clearpage

\begin{table} 
\caption{Dwarf Nova Parameters}
\begin{tabular}{llllllcccc} 
%\multicolumn{10}{c}
\hline \\   
System  & $P_{orb}$ &$t_{rec}$&V$_{q}$  &V$_{o}$  &  Sec.   & i       &  $M_{wd}$   &$E_{B-V}$& Distance \\ 
Name    & (days)    & (days)  &         &         &         & (deg)   &$(M_{\odot})$&         & (pc) \\ \hline 
BV Cen  & 0.610108 &   150   &  12.6   &  10.5   &   G5-8V &$62 \pm5$&$0.83\pm 0.1$ & 0.10    & 500    \\ 
CH UMa  & 0.343    &   204   &  15.3   &  10.7   &   K4-M0V&$21 \pm4$&  &         & 300    \\ 
UU Aql  & 0.14049  &   71    &  16.0   &  11.0   &   M2-4V &         &              &         & 150-350 \\ 
\hline 
\end{tabular} 
\end{table} 

\begin{deluxetable}{lccccccc} 
\tablecaption{{\it{FUSE}} Observations} 
\tablewidth{0pc} 
%\tablenum{2}                                                                               
\tablehead{                                   
\colhead{Target}              & 
\colhead{Data ID}             & 
\colhead{Aperture}            & 
\multicolumn{2}{c}{Time of Observation} & 
\colhead{t$_{exp}$}           &
\colhead{$\lambda_{centered}$} &
\colhead{S/N(0.1\AA)}   
} 
\startdata        
UU Aql  &  C1100301000 &  LWRS 30x30 &  2004-05-16 & 13h:48m:00s &  16,121s  & 1035\AA   & 5.15 \\ 
BV Cen  &  D1450301000 &  LWRS 30x30 &  2003-04-13 & 20h:26m:00s &  26,545s  & 1035\AA   & 5.9 \\ 
CH UMa  &  D1450201000 &  LWRS 30x30 &  2003-04-02 & 22h:00m:00s &  17,311s  & 1035\AA   & 3.6  \\ 
\enddata 
\end{deluxetable}    

\begin{deluxetable}{lcccccc}   
\tablecaption{{\it{IUE}} Observations} 
%\tablenum{3}                                                                               
\tablehead{                                    
\colhead{Target}& 
\colhead{SWP No. }& 
\colhead{Aperture}& 
\colhead{Dispersion}& 
\colhead{Date}& 
\colhead{Time 0f Mid-Exposure}& 
\colhead{t$_{exp}$(s)} 
} 
\startdata         
BV Cen   & 26623  &       Lg   &     Low  &     08/16/85&       17:17:00&             14,400\\ 
CH UMa   & 56270 &        Lg &       Low &      12/06/95&       12:15:49&             13,800\\ 
\enddata 
\end{deluxetable} 

\clearpage 

\begin{deluxetable}{ll}   
\tablecaption{Masked Spectral Regions} 
%\tablenum{4}                                                                                
\tablehead{                                    
\colhead{Target}&\colhead{Emission Features } 
} 
\startdata                                                 
UU Aql & for             $ \lambda <$ 1050\AA\ if $F_{\lambda}>6   \times 10^{-15}$ergs$~$s$^{-1}$cm$^{-2}$\AA$^{-1}$ \\ 
      & for 1050 \AA\ $ < \lambda <$ 1185\AA\ if $F_{\lambda}>1.8 \times 10^{-14}$ergs$~$s$^{-1}$cm$^{-2}$\AA$^{-1}$  \\ 
      & 975-985\AA  \\ 

CH UMa & for $\lambda < 957$\AA\ if $F_{\lambda} > 1 \times 10^{-14}$ergs$~$s$^{-1}$cm$^{-2}$\AA$^{-1}$   \\ 
      & 970-980, 987-994, 1023-1042, 1071-1090, 1107-1136, 1167-1180, 1200-1260, \\   
      & 1285-1315, 1380-420, 1532-1560, 1630-1650, 1846-1940\AA \\ 

BV Cen &  for $\lambda < 1185$\AA\ if $F_{\lambda} > 2.5 \times 10^{-14}$ergs$~$s$^{-1}$cm$^{-2}$\AA$^{-1}$  \\ 
      & 1190-1220, 1320-1340, 1375-1415, 1530-1570, 1625-1650, 1835-1880\AA \\ 

\enddata 
\end{deluxetable} 

\begin{deluxetable}{llclccccc} 
\tablecaption{UU Aql, BV Cen, and CH UMa Fitting Results} 
\tablehead{
\colhead{System}      & 
\colhead{Spectrum}    & 
\colhead{d}           & 
\colhead{$\chi^{2}$}  & 
\colhead{$T_{wd}$}    & 
\colhead{$T_{belt}$}  & 
\colhead{$\dot{M}$}   & 
\colhead{\% Flux}     &
\colhead{Fig}            \\
\colhead{  }   & 
\colhead{  }   & 
\colhead{(pc)} & 
\colhead{  }   & 
\colhead{($103$K)}  & 
\colhead{($103$K)}  & 
\colhead{($M_{\sun}$/yr)}   & 
\colhead{(Contribution)}    &
\colhead{ }  
} 
\startdata 
UU Aql & {\it{FUSE}}      & 350 & 0.96  & 27 & -  &                 -   & 100(WD)        &  4  \\ 
       & {\it{FUSE}}      & 150 & 1.13  & 17 & -  &                 -   & 100(WD)        &  -  \\ 
       & {\it{FUSE}}      & 350 & 1.13  & -  & -  & $5 \times 10^{-11}$ & 100(Disk)      &  5  \\ 
       & {\it{FUSE}}      & 150 & 1.24  & -  & -  & $1 \times 10^{-11}$ & 100(Disk)      &  -  \\ 
       & {\it{FUSE}}      & 350 & 0.714 & 24 & -  & $2 \times 10^{-11}$ & 73(WD)/27(Disk) & - \\ 
       & {\it{FUSE}}      & 150 & 1.13  & 17 & -  & $3 \times 10^{-11}$ & 60(WD)/40(Disk) & - \\ 
       & {\it{FUSE}}      & 350 & 0.738 & 24 & 33 &             -       & 57(WD)/43(Belt) & 6 \\ 
       & {\it{FUSE}}      & 150 & 1.021 & 16 & 29 &             -       & 77(WD)/23(Belt) & - \\ 
CH UMa & {\it{FUSE}}      & 314 & 0.184 & 40 & -  &     -               & 100(WD)        &  7  \\ 
       & {\it{FUSE}}      & 313 & 0.213 &  - & -  & $5 \times 10^{-12}$ & 100(Disk)      &  -  \\ 
       & {\it{FUSE}}+{\it{IUE}} & 300 & 3.02  & 31 & -  &     -               & 100(WD)        &  -  \\ 
       & {\it{FUSE}}+{\it{IUE}} & 300 & 2.02  & -  & -  & $3 \times 10^{-9}$  & 100(Disk)      &  -  \\ 
       & {\it{FUSE}}+{\it{IUE}} & 300 & 1.87  & 22 & -  &$6.4\times 10^{-11}$ & 21(WD)/79(Disk)&  -  \\  
       & {\it{FUSE}}+{\it{IUE}} & 300 & 2.00  & 26 & 50 &      -              & 77(WD)/23(Belt)&  -  \\    
BV Cen & {\it{FUSE}}      & 435 & 0.27  & 40 & -  &             -       & 100(WD)        &  8  \\ 
       & {\it{FUSE}}      & 1255& 0.27  & -  & -  & $3 \times 10^{-9}$  & 100(Disk)      &  -     
\enddata 
\end{deluxetable} 

\clearpage 

\begin{table}
\caption{Surface Temperatures of White Dwarfs in Dwarf Novae above the Period Gap}
\begin{tabular}{lccl} 
%\multicolumn{4}{c} 
\hline 
\hline 
System Name &  Period & $T_{eff}$ &     References   \\            
       &   (min)  &  (Kelvin)     &                 \\ \hline 
BV Cen  &   878.6  &   40,000:    &      This paper           \\ 
RU Peg  &   539.4  &   49,000     &      \citet{sio04}        \\ 
Z Cam   &   417.4  &   57,000     &      \citet{har05}        \\ 
RX And  &   302.2  &   34,000     &      \citet{sio01}        \\ 
SS Aur  &   263.2  &   31,000     &      \citet{sio04}        \\ 
U Gem   &   254.7  &   31,000     &      \citet{sio98}        \\ 
WW Ceti &   253.1  &   26,000     &      \citet{god06}        \\ 
UU Aql  &   202.3  &   27,000:    &      This paper           \\ 
\hline 
\end{tabular} 
\end{table} 

\clearpage 

Figure Captions 

\figcaption{The flux $F_{\lambda}$ (ergs/cm$^{2}$/s/\AA) versus 
wavelength (\AA) {\it{FUSE}}
spectrum of the U Gem-type dwarf nova UU Aql during quiescence. The 
identifications of the strongest neutral and ionized line features as well 
as rotational-vibrational transitions of molecular 
hydrogen are marked with vertical 
tick marks. The molecular hydrogen absorption are modestly affecting 
the continuum, slicing it at almost regular wavelength intervals.   
We identify the most prominent molecular hydrogen 
absorption lines by their band 
(Werner or Lyman), upper vibrational level (1-16), and rotational transition 
(R,P, or Q) with lower rotational state (J=1,2,3). 
Air glow line features are indicated with the earth 
symbol. The C\,{\sc iii} (977\AA ) and O\,{\sc vi} doublet sharp emissions 
are due to heliocoronal emission often present 
in some of the {\it{FUSE}} channels when the target (i.e. UU Aql) is faint.} 

\figcaption{The flux $F_{\lambda}$ (ergs/cm$^{2}$/s/\AA) versus 
wavelength (\AA) {\it{FUSE}}  
spectrum of the U Gem-type dwarf nova BV Cen during quiescence. The 
identifications of the strongest neutral and ionized line features as well 
as rotational-vibrational transitions of molecular 
hydrogen are marked with vertical 
tick marks. Air glow line features are indicated with the earth 
symbol. The O\,{\sc i} emissions are also due to air glow. 
The broad C\,{\sc iii} (977\AA ) 
and O\,{\sc vi} (doublet) features might 
be real broad emissions from the source, as these 
correspond to the left wing of the Lyman $\gamma$ 
\& $\beta$ (respectively) and not 
much flux is expected there.  The N\,{\sc i} and N\,{\sc ii} 
sharp emission features are geocoronal.}   

\figcaption{The flux F$\lambda$ (ergs/cm$^{2}$/s/\AA) versus wavelength 
(\AA) {\it{FUSE}} 
spectrum of the U Gem-type dwarf nova CH UMa during quiescence. The 
identifications of the strongest neutral and ionized line features as well 
as rotational-vibrational transitions of molecular hydrogen are marked
with vertical tick marks. Several terrestrial 
line features are indicated with the earth 
symbol. There are some broad emission features associated with the source: 
we identify here the O\,{\sc vi} doublet, C\,{\sc iii} 
(at 977\AA\ and 1175\AA ). 
There could be some N\,{\sc iv} emission in the short 
wavelengths (between 920\AA\ 
and 925\AA ), but the increase of flux in the very short wavelengths 
suggests that most the flux at $\lambda <930$\AA\ could be 
higher orders of the hydrogen Lyman series associated with air glow       
contamination and possibly some detector noise.}    

\figcaption{A single temperature WD fit to the {\it{FUSE}} spectrum of UU Aql 
for a distance of 350 pc. The best-fit yielded T$_{eff} = 27,000$K
$\pm3000$K, log $g = 9$. The 
white dwarf contributes 100\% of the FUV flux.} 

\figcaption{An accretion disk-only fit to the 
{\it{FUSE}} spectrum of UU Aql for 
a distance of 350 pc. The best-fit corresponded to M$_{wd} = 
0.8 M_{\sun}$, $i = 41$ degrees, $\dot{M} = 5 \times 10^{-11} 
M_{\sun}$/yr. The accretion disk contributes 100 \% of the FUV flux.}

\figcaption{A WD + belt fit to the {\it{FUSE}} spectrum of UU Aql for 
a distance of 350 pc. The dotted line is the white dwarf flux 
component, the dashed line is the belt flux component and the solid 
line is the combined two-temperature fit.
This best fit two temperature WD yields $\chi^{2} = 0.74$ 
with the cooler white dwarf portion ( T$_{eff} = 24,000$K) giving 57\% of 
the flux and the hotter belt (T$_{belt} = 33,000$K) providing 43\%.} 

\figcaption{A single temperature WD fit to the {\it{FUSE}} spectrum 
of CH UMa. The 
best-fit yielded T$_{eff} = 40,000\pm1,000$K, log $g = 9.5$
and a projected rotational velocity of $V_{rot} \sin{i}=200$km/s. 
The portions of the spectrum that have been masked are shown in blue and 
include emission lines (either intrinsic to the source or due to air glow) and 
ISM molecular hydrogen absorption lines (which have been matched using a simple
ISM model to identify the exact location of the ISM lines). Therefore
the fit is between the model and the red portions of the observed
{\it{FUSE}} spectrum. The absence of strong absorption features in the
lower panel were modeled assuming low Si and C abundances and led to
a lower $\chi^2$ than the same model with solar C and Si abundances but with
a high projected rotational velocity ($\approx 1,000$km/s).}  

\figcaption{A single temperature WD fit to the 
{\it{FUSE}} spectrum of BV Cen. The 
best-fit yielded T$_{eff} = 40,000$K$\pm1000$K, a projected rotational
velocity of 500km/s, and d=435pc, assuming $M=0.83M_{\odot}$.}

%\figcaption{A WD plus accretion disk fit to 
%the {\it{FUSE}} plus {\it{IUE}} spectrum of UU 
%Aql for d = 350 pc. The white dwarf T$_{eff} = 24,000$K$\pm 3000$K, 
%$\dot{M} = 2 \times 10^{-11} M_{\sun}$/yr, for $i = 41$ degrees, log $g = 
%8.2$, with the WD contributing 73\% of the FUV flux and the accretion disk 
%providing 27\% of the FUV flux.}
%
%\figcaption{A WD plus accretion disk fit to 
the {\it{FUSE}} plus {\it{IUE}} spectrum of UU 
%Aql for d = 350 pc. The white dwarf T$_{eff} = 24,000$K$\pm 3000$K, 
%$\dot{M} = 2 \times 10^{-11} M_{\sun}$/yr, for $i = 41$ degrees, log $g = 
%8.2$, with the WD contributing 73\% of the FUV flux and the accretion disk 
%providing 27\% of the FUV flux.}
%

\clearpage 
\begin{figure} 
\plotone{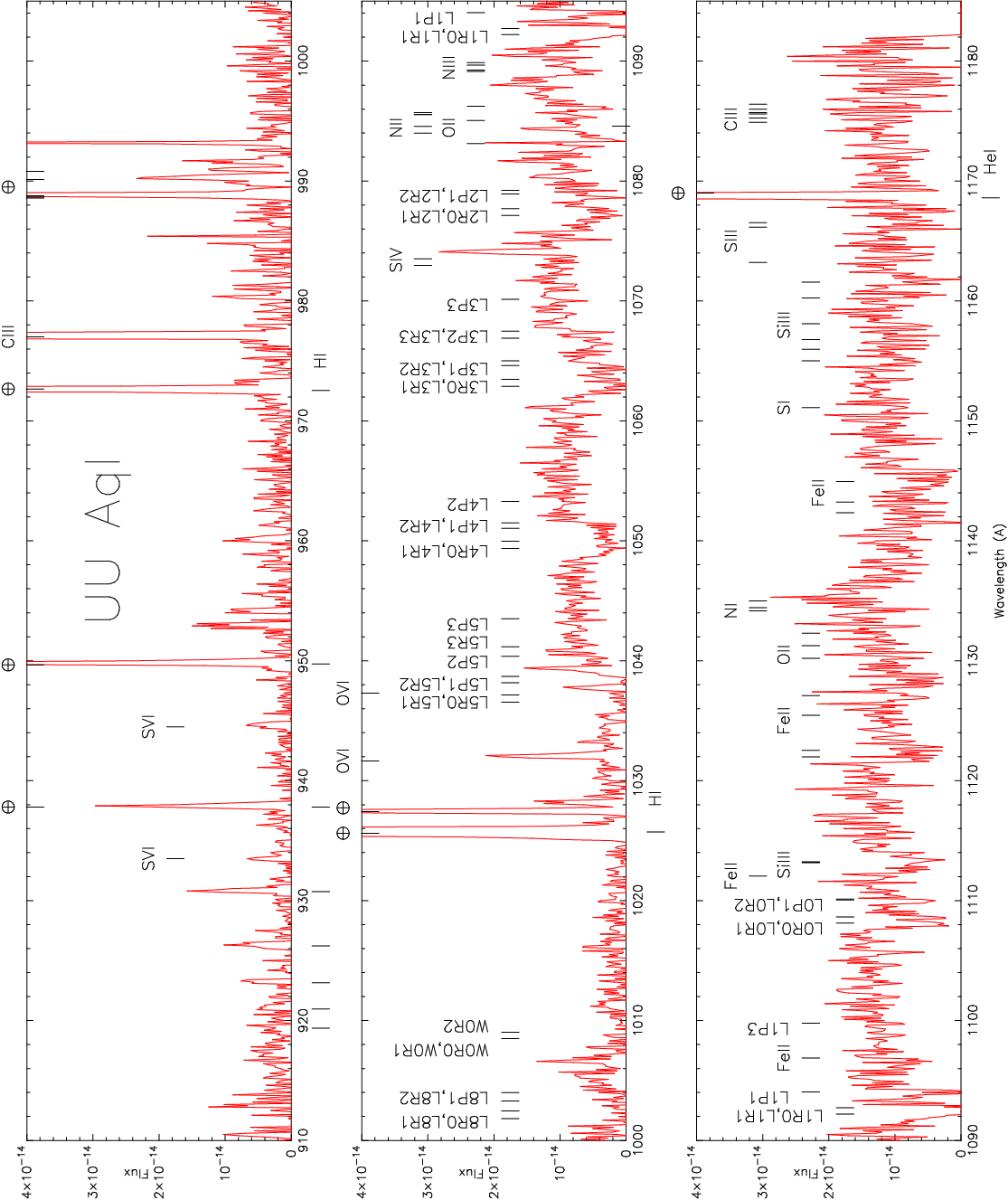} 
\end{figure}

\clearpage 
\begin{figure} 
\plotone{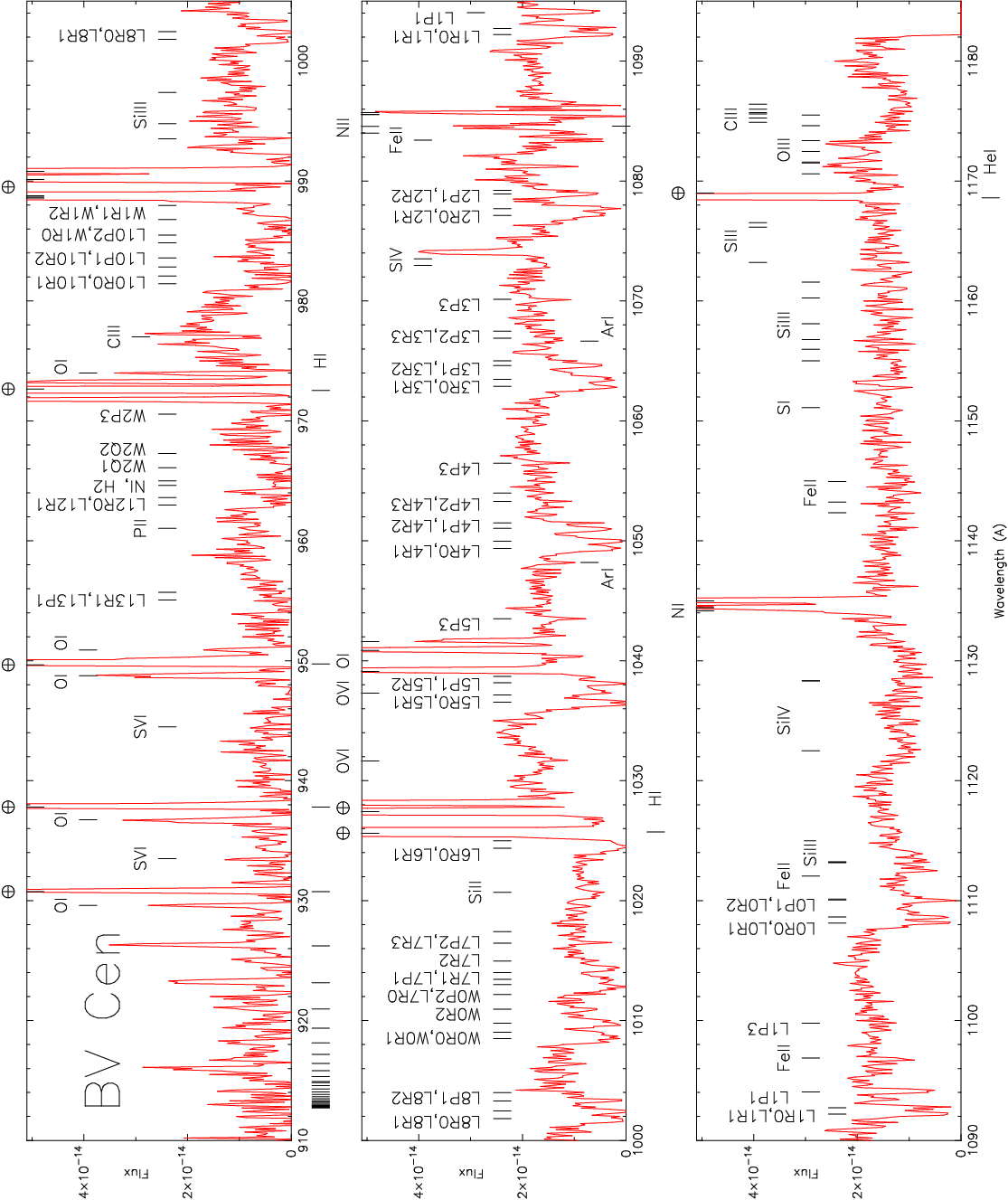} 
\end{figure}

\clearpage 
\begin{figure} 
\plotone{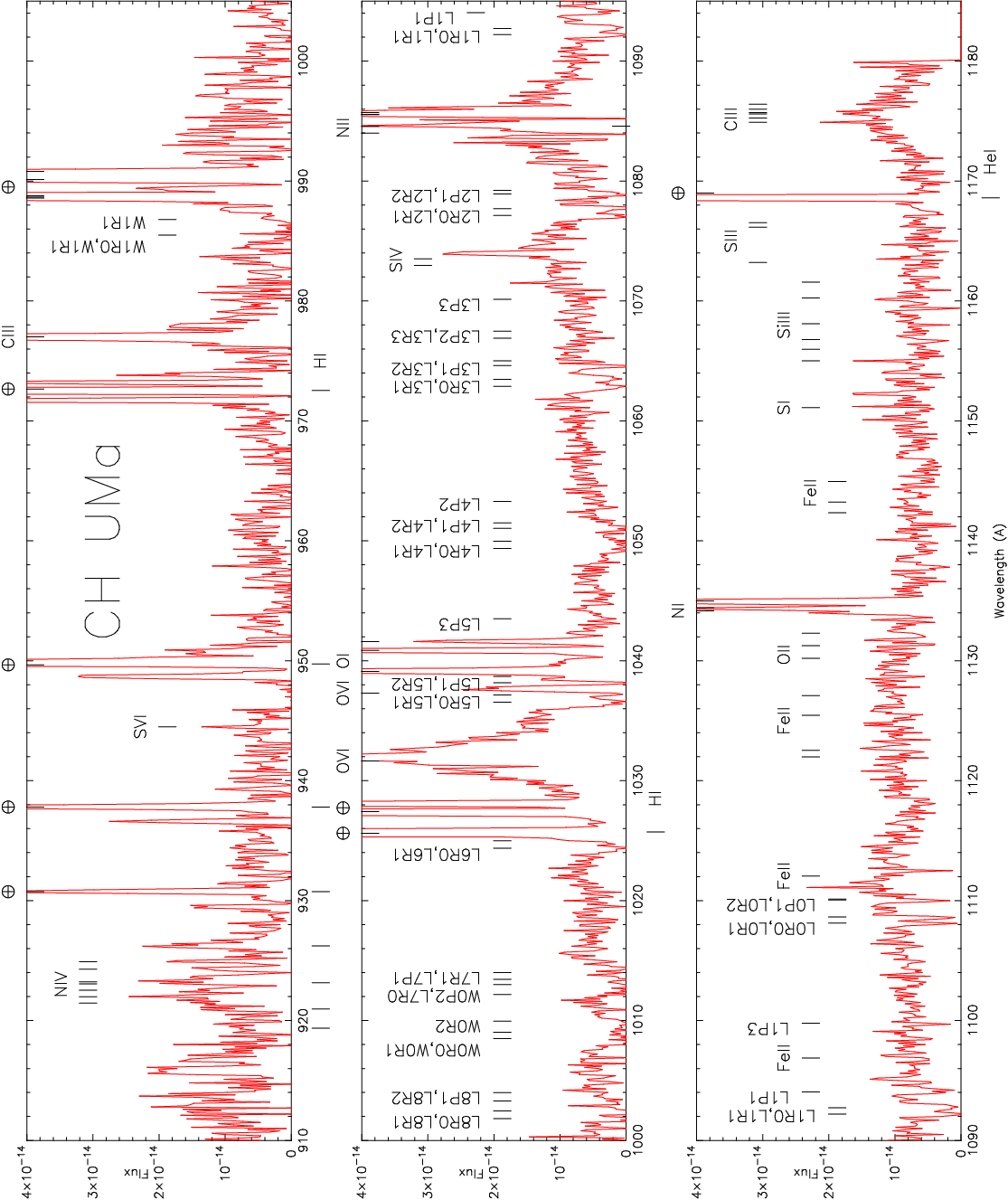} 
\end{figure}

\clearpage 
\begin{figure} 
\plotone{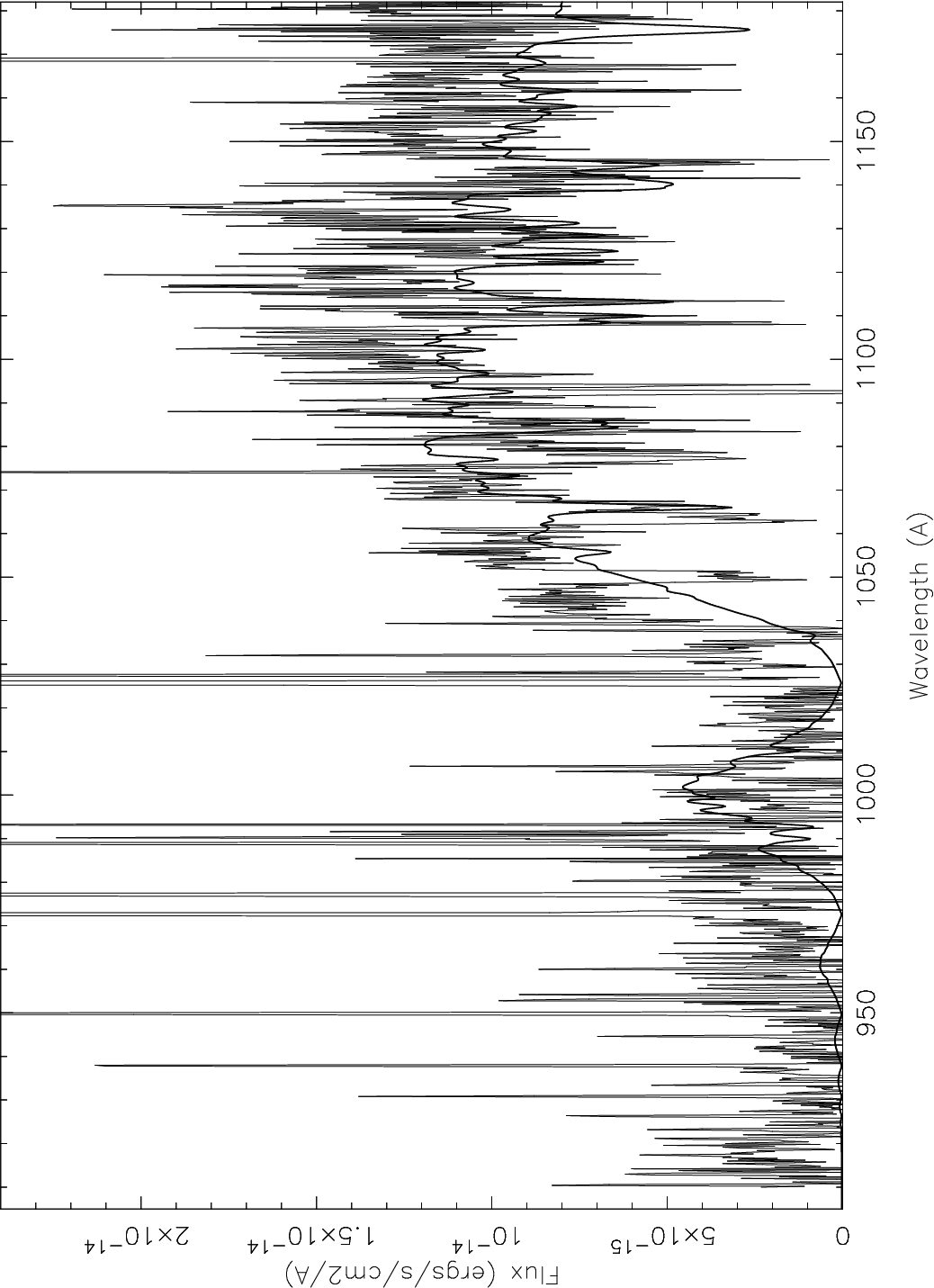} 
\end{figure}

\clearpage 
\begin{figure} 
\plotone{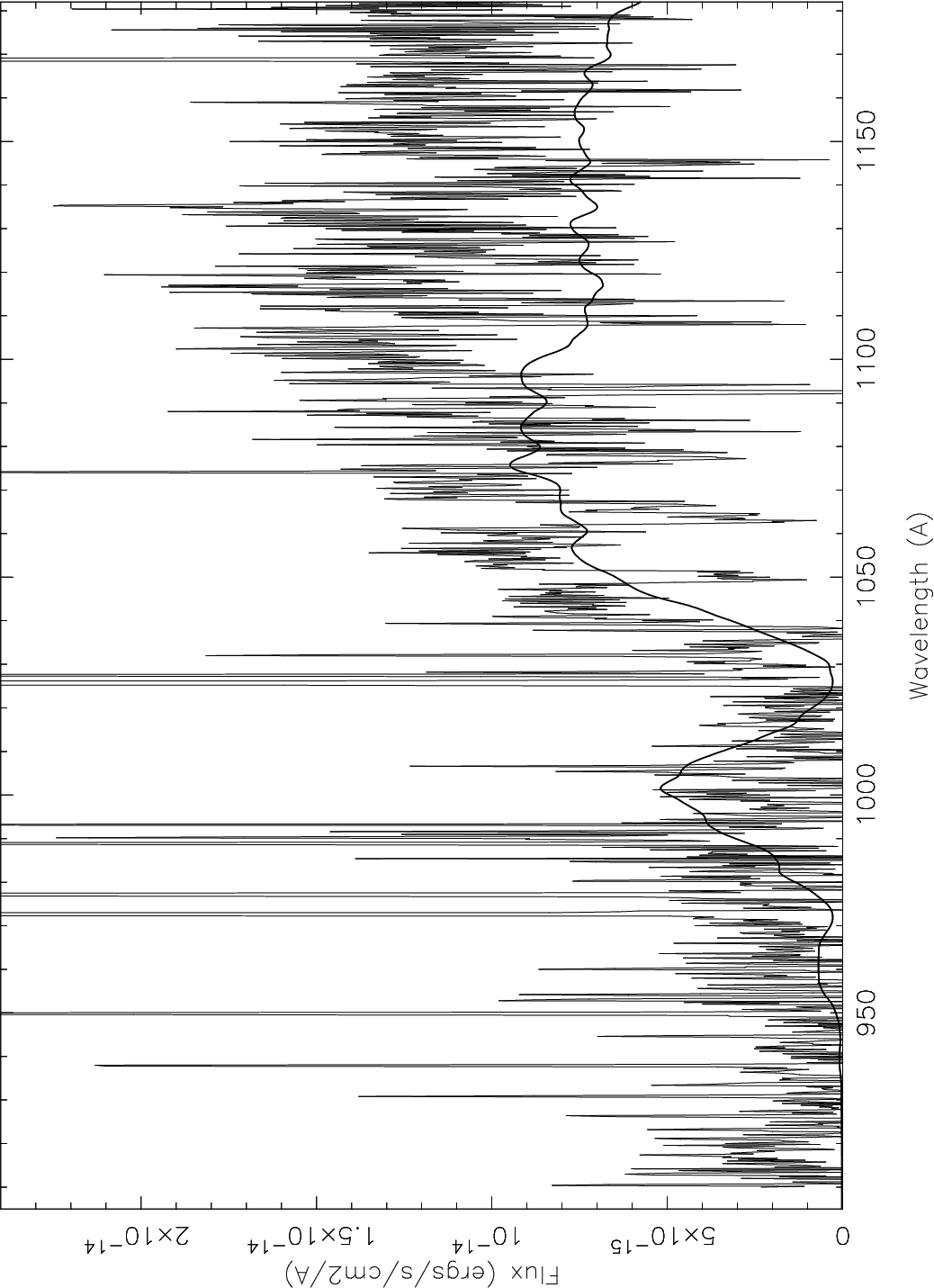} 
\end{figure} 

\clearpage 
\begin{figure} 
\plotone{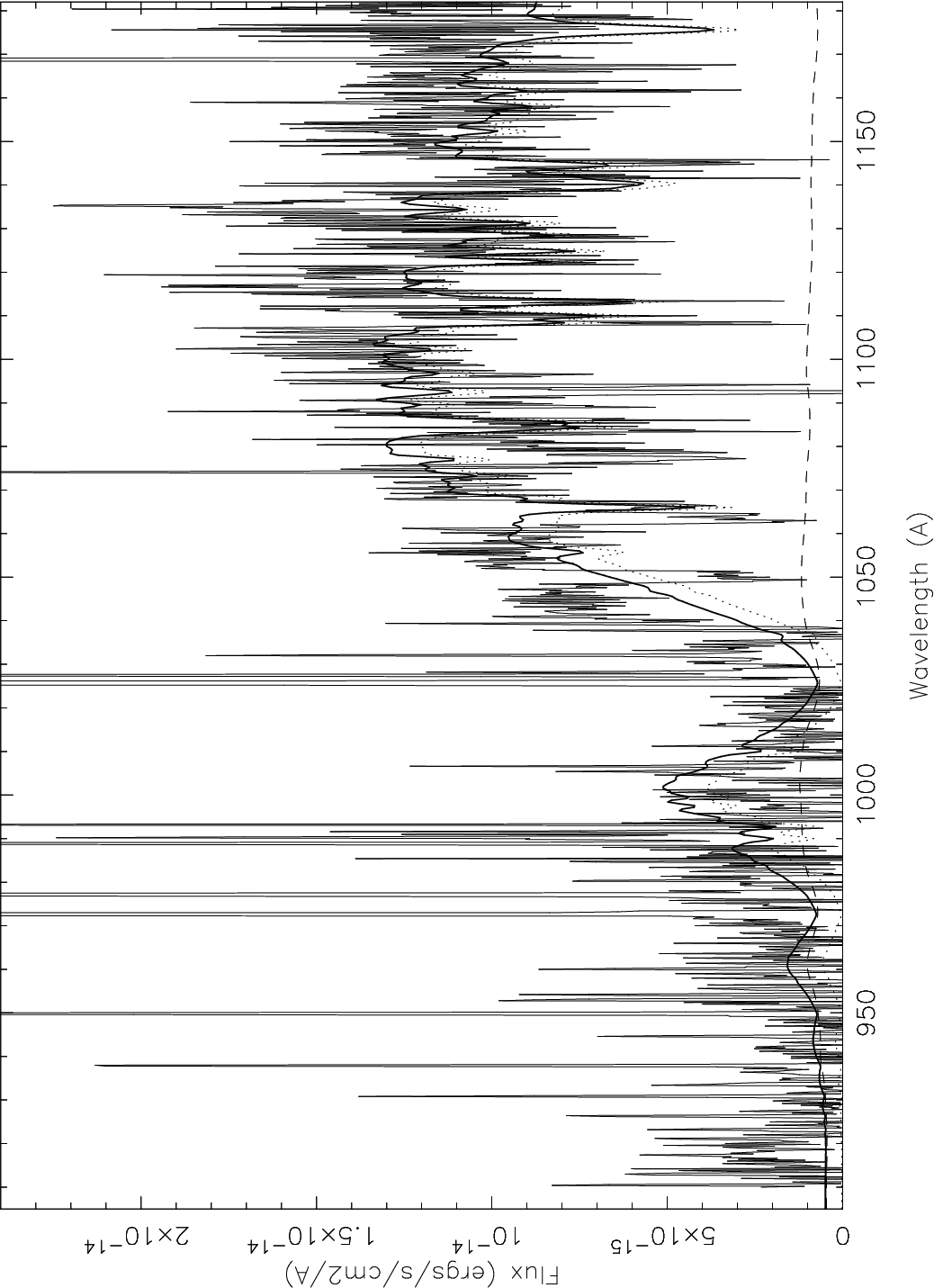} 
\end{figure}

\clearpage 
\begin{figure} 
\plotone{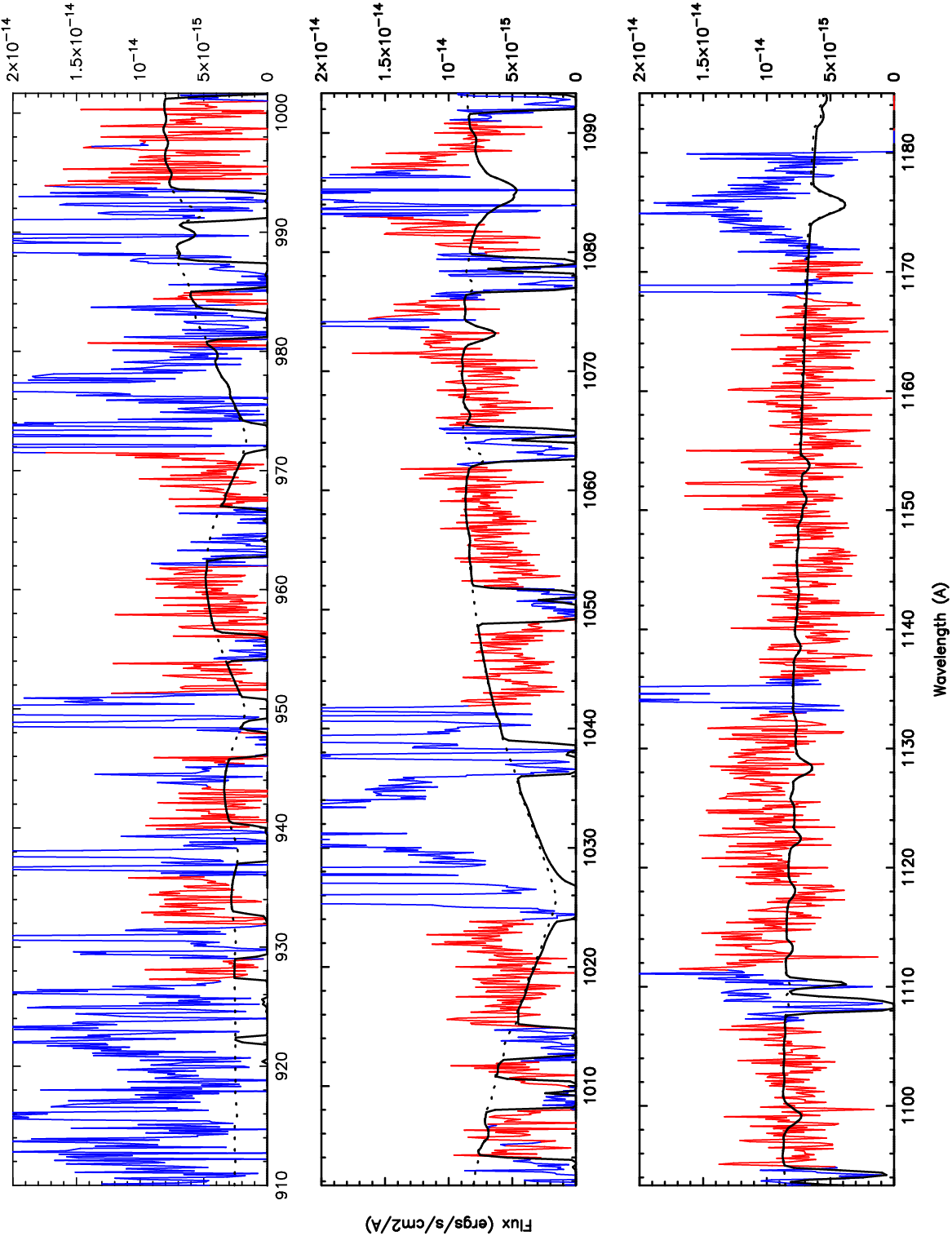}  
\end{figure}

\clearpage 
\begin{figure} 
\plotone{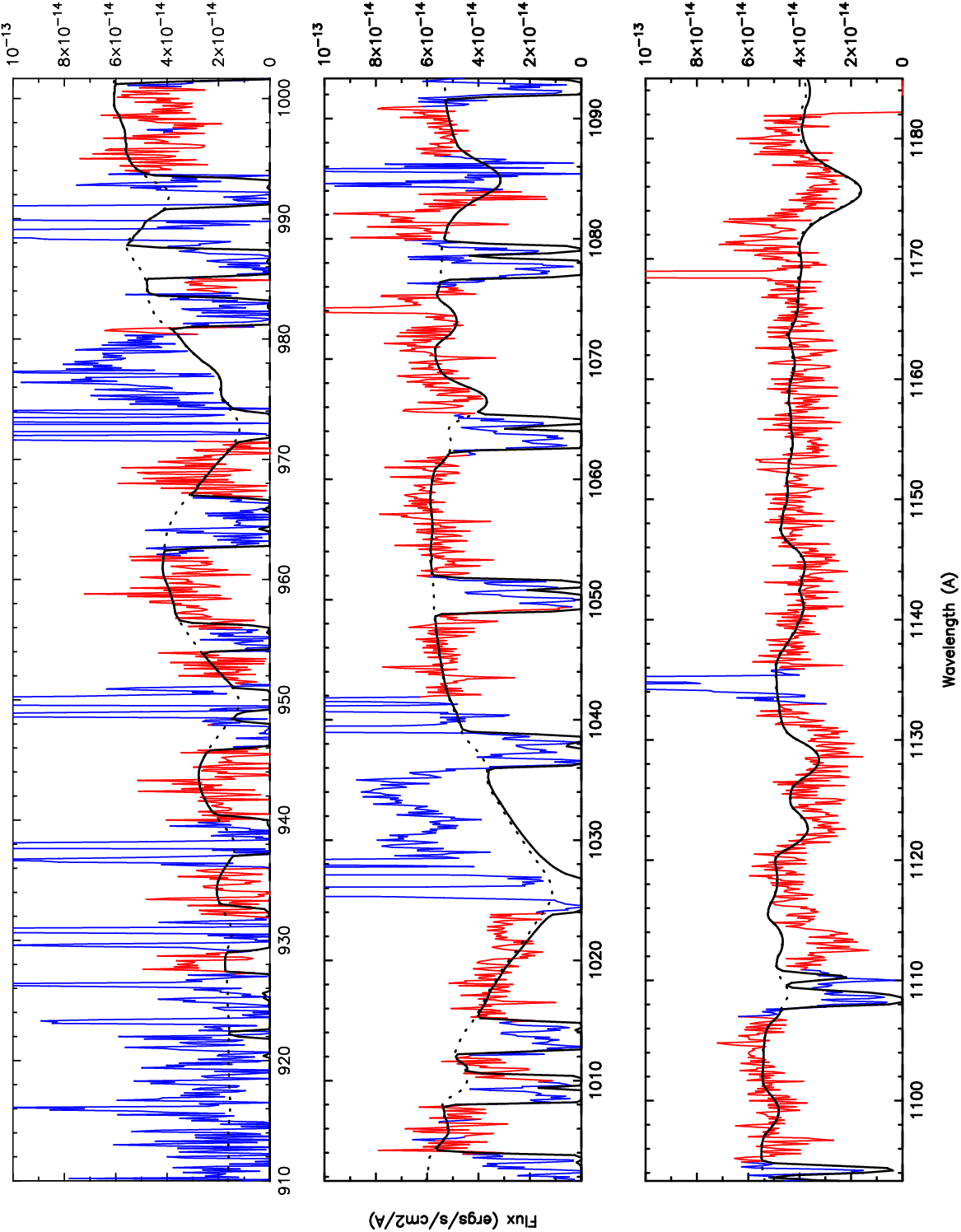}
\end{figure}

\end{document}